\documentclass[12pt, draftclsnofoot, onecolumn]{IEEEtran}
\pdfoutput=1

\usepackage{lipsum}
\usepackage{graphicx,amssymb,lineno}
\usepackage{amsmath,amsfonts,amssymb}
\usepackage{array}
\usepackage{mathrsfs}
\newtheorem{theorem}{Theorem}

\usepackage{algorithm}
\usepackage[usenames]{color}
\usepackage{float}
\usepackage{graphics}
\usepackage{booktabs}
\usepackage{multicol}
\usepackage{multirow}
\usepackage[noadjust]{cite}
\usepackage{stfloats}
\usepackage{algpseudocode}
\usepackage{graphicx,graphics,color,epsfig,subfigure,graphpap,rotate}
\usepackage{times, verbatim, subfigure, epsfig, graphicx, latexsym, amsmath}
\usepackage{url}
\usepackage{subfigure}
 \usepackage{indentfirst}

\newtheorem{remark}{Remark}
\newtheorem{lemma}[theorem]{Lemma}

\renewcommand{\algorithmicensure}{\textbf{Iteration:}}

\renewcommand{\algorithmicensure}{\textbf{Output:}}
\DeclareMathOperator*{\tr}{tr}
\DeclareMathOperator*{\st}{s.t.}
\newcommand\Ab{\ensuremath{{\boldsymbol A}}}
\newcommand\Bb{\boldsymbol B}
\newcommand\Cb{\ensuremath{{\boldsymbol C}}}
\newcommand\Dbb{\ensuremath{{\boldsymbol D}}}

\newcommand\Hb{\ensuremath{{\boldsymbol H}}}
\newcommand\Wb{\ensuremath{{\boldsymbol W}}}
\newcommand\Ib{\ensuremath{{\boldsymbol I}}}

\newcommand\Fb{\ensuremath{{\boldsymbol F}}}

\newcommand\Kb{\ensuremath{{\boldsymbol K}}}

\newcommand\Xb{\ensuremath{{\boldsymbol X}}}

\newcommand{\zero}{{\bf{0}}}

\newcommand\cb{\ensuremath{{\boldsymbol c}}}
\newcommand\db{\ensuremath{{\boldsymbol d}}}

\newcommand\fb{\ensuremath{{\boldsymbol f}}}
\newcommand\kb{\ensuremath{{\boldsymbol k}}}

\newcommand\qb{\ensuremath{{\boldsymbol q}}}
\newcommand\ssb{\ensuremath{{\boldsymbol s}}}

\newcommand\ub{\ensuremath{{\boldsymbol u}}}
\newcommand\wb{\ensuremath{{\boldsymbol w}}}
\newcommand\rb{\ensuremath{{\boldsymbol r}}}
\newcommand\xb{\ensuremath{{\boldsymbol x}}}
\newcommand\yb{\ensuremath{{\boldsymbol y}}}

\newcommand\mtheta{\ensuremath{{\boldsymbol \theta}}}

\newcommand\mpsi{\boldsymbol{\mathnormal \psi}}

\newcommand{\mXi}{{\boldsymbol{\mathnormal\Xi}}}

\newcommand{\mPhi}{{\boldsymbol{\mathnormal\Phi}}}
\newcommand{\mphi}{{\boldsymbol{\mathnormal\phi}}}

\renewcommand\vec {\text{vec}}

\newcommand{\Rmnum}[1]{\expandafter\@slowromancap\romannumeral #1@}

\makeatother

\def\BibTeX{{\rm B\kern-.05em{\sc i\kern-.025em b}\kern-.08em
    T\kern-.1667em\lower.7ex\hbox{E}\kern-.125emX}}
\begin{document}

\title{Reconfigurable-intelligent-surface-assisted Downlink Transmission Design via Bayesian Optimization}


\author{Dong Wang, Xiaodong Wang,~\IEEEmembership{Fellow,~IEEE},  
 Fanggang Wang,~\IEEEmembership{Senior Member,~IEEE}

\thanks{D. Wang and F. Wang are with the State Key Laboratory of Rail Traffic Control and Safety, Beijing Jiaotong University, Beijing
100044, China (e-mail: \{17111034, wangfg\}@bjtu.edu.cn).

X. Wang is with the Department of Electrical Engineering, Columbia
University, New York, NY 10027 USA (e-mail: wangx@ee.columbia.edu).
}
}

\maketitle

\begin{abstract}
This paper investigates the transmission design in the  reconfigurable-intelligent-surface (RIS)-assisted downlink system. The channel state information (CSI) is usually difficult to be estimated at the base station (BS) when the RIS is not equipped with radio frequency chains. 
In this paper, we propose a downlink transmission framework with unknown CSI via Bayesian optimization. Since the CSI is not available at the BS, we treat the unknown objective function as the black-box function and take the  beamformer, the phase shift, and the receiving filter as the input. Then the objective function is decomposed as the sum of low-dimension subfunctions to reduce the complexity. By re-expressing the power constraint of the BS in spherical coordinates, the original constraint problem is converted into an equivalent unconstrained problem. The users estimate the sum MSE of the training symbols as the objective value and feed it back to the BS. We assume a Gaussian prior of the feedback samples and the next query point is updated by minimizing the constructed acquisition function. Furthermore, this framework can also be applied to the power transfer system and fairness problems. Simulation results validate  the effectiveness of the proposed transmission scheme in the downlink data transmission and power transfer.
\end{abstract}
\section{Introduction}\label{intro}
The deployment of the massive multiple-input multiple-output (MIMO), millimeter wave are expected to achieve the high data rate and massive device connections in the wireless networks \cite{future_net}. However, the dense deployments of multi-antenna base stations (BSs) and access points (APs) have raised serious concerns on their energy consumption, which the wireless network operators struggle to support millions of users, especially the users in the unfavorable propagation environments \cite{Renzo19}. The reconfigurable intelligent surface (RIS) is regarded as a promising technology in the future sixth-generation (6G) mobile communications network due to its passive and low-cost characteristics \cite{toward_future,turtorial,swipt,tccn}. A RIS is a meta-surface equipped with integrated electronic circuits that can be programmed to alter an incoming electromagnetic field in a customizable way. By suitably adjusting the phase shift of each reflecting element, the RIS consumes much less energy than a regular amplify-and-forward relay \cite{gong2019smart}. Furthermore, the RIS can be flexibly deployed at the building facades, indoor walls, and ceilings, and applied to various communication scenarios such as the cellular networks, the wireless power transfer networks, and the unmanned aerial vehicle networks, etc.

The existing works on the RIS-assisted communication system mainly focus on cellular scenarios. The authors in \cite{MISO_wuGC} investigated the point-to-point multi-input
single-output (MISO) wireless system and the total received signal power is maximized based on the semidefinite relaxation (SDR). In order to reduce the high computational complexity incurred by SDR, the authors in \cite{MISO_yuICCC} proposed low complexity algorithms by majorization-minimization (MM) and manifold optimization. In \cite{Practical_MISO}, the authors analyzed the practical model of the reflection coefficient and applied it to the RIS-assisted wideband orthogonal frequency-division multiplexing (OFDM) system. The authors in \cite{OFDM_MISO} maximized the achievable rate in the RIS-assisted OFDM system.  The authors in \cite{TWC_MU_MISO} proposed the alternating optimization algorithm and the two-stage algorithm to minimize the total transmit power at the AP in the multiple-user MISO system, while in \cite{MU_TWC}, the authors developed  the energy-efficient designs based on the popular gradient descent search and the sequential fractional programming. A simplified semidefinite programming-based reflecting scheme and a simplified maximizing the minimum Euclidean distance precoding design were studied in the RIS-assisted MIMO system. Moreover, the authors in \cite{mutigroup_MISO} proposed two transmission schemes to maximize the sum rate  based on the MM in the multigroup multicast communication systems.

The aforementioned transmission designs assume the channel state information (CSI) is available at the BS and RIS, which means that the channel estimation is required before the beamforming and the phase shift design. The channel estimation methods in the RIS-assisted system can be categorized into two types. One is that the radio frequency (RF) chains are assembled  in the RIS for receiving the training pilots at each element.  The authors in \cite{hardware_est} presented an alternating optimization approach for explicit estimation of the channel gains at the RIS elements attached to the single RF chain. Another approach was to select the RIS reflection matrix from quantized codebooks via beam training, but the complexity and the overhead increase rapidly when the number of the elements goes larger \cite{deep_est}. However, the equipped RF-chains of the RIS cause huge energy consumption and hardware cost.  The authors in \cite{luo} adopted the spatial modulation to reduce the number of RF-chains. The other type is that the cascaded channel estimation without equipping RF-chains at the RIS.  In \cite{est_onebyone}, the authors switched the RIS elements ON one-by-one and estimated the RIS-assisted channels at the BS based on received pilot symbols from the users.  The BS-RIS-user cascaded channel was estimated based on a two-stage algorithm that includes a sparse matrix factorization stage and a matrix completion stage \cite{He2020CascadedCE}. A two-timescale channel estimation framework was proposed in \cite{two_timescale} to exploit the property of the BS-RIS and RIS-user channels. Moreover, the authors in \cite{matrix_channel} proposed a matrix-calibration-based cascaded channel estimation for the RIS-assisted multi-user MIMO. The authors in \cite{adaptive_ris} proposed an adaptive transmission protocol in conjunction with a progressive channel estimation method for the wideband OFDM system.  It is worth noting that the cascade channel estimation is usually adopted in the uplink system. The channel estimation for the downlink system is difficult to operate, which remains for future research. Although the transmission design without the CSI has been studied in the wireless power transfer and the RF identification scenarios \cite{energy_bm,iot_blind}, to the best our knowledge, there is no literature that investigated the transmission design in the downlink RIS-assisted system without the CSI.
Therefore, it is crucial to design a downlink transmission scheme without the CSI and keep the RIS passive or nearly passive.

In this paper, we propose a transmission framework with unknown CSI for downlink RIS-assisted system via Bayesian optimization. Bayesian optimization is a sequential design strategy for global optimization of black-box functions, which has emerged as a powerful solution for these varied design problems such as interactive user interface, information extraction, automatic machine learning, and sensor network, etc \cite{BO_take,brochu2010tutorial,inproceedings}. The Bayesian optimization adds a Bayesian methodology to the iterative optimizer paradigm by incorporating a prior model on the space of possible target functions. A simpler surrogate model of the objective function which is cheaper to evaluate and will be used instead to solve the optimization problem. The main contributions of our work are summarized as follows:

\setlength{\parindent}{2em}
\begin{itemize}
 \item To the best of our knowledge, this paper is the first attempt to focus on the downlink transmission scheme without the channel estimation. Motivated by the Bayesian optimization, we formulate the optimization problems to minimize the mean square error (MSE) in the data transmission. Since the CSI of the BS-RIS and the RIS-users are not available at the BS and the RIS, this optimization cannot be solved analytically. By introducing the Bayesian optimization, the sum MSE can be regarded as the objective function. The input of the objective function is the beamformer at the BS, the phase shift at the RIS, and the receiver filter at the users. Then, we reformulate the original optimization problem to an unconstraint problem by using the spherical coordinates. It is worth noting that the dimension of the input vector is high when the number of antennas of the BS and the number of elements of the RIS is large. We deal with this challenge by treating the objective function additive function of mutually exclusive lower-dimensional components \cite{additive_bo}.
  \item We propose a framework of the  transmission scheme in the RIS-assisted downlink system. The proposed downlink data transmission scheme can be easily extended to the MIMO system, the multi-user MIMO system. In addition, by considering the fairness among the users, the maximum of individual MSE can be minimized by the Bayesian optimization.  Furthermore, this setup of our work can also apply to the wireless power transfer system by choosing the received signal strength as the objective function.
  \item Simulation results show that the proposed downlink data transmission scheme based on the Bayesian optimization can effectively reduce the sum MSE. For the downlink wireless power transfer, the proposed transmission scheme can improve the received power and achieve acceptable performance compared with the known CSI scheme. We also validate the performance by considering the fairness of users. Moreover, the proposed transmission scheme is robust to the slow fading channel, where the updated variable is dynamically learned by each feedback.
\end{itemize}

The remainder of the paper is organized as follows: Section \ref{preli} introduce the background of the Bayesian optimization. Section \ref{Sys} introduces the system model and problem formulation in the downlink data transmission system and the wireless power transfer system, respectively. The proposed downlink transmission scheme via Bayesian optimization is studied in Section \ref{BO_app}. Then, Section \ref{simula} provides simulation results. Finally, we conclude this paper in Section \ref{S7}.

\emph{Notation}: Boldface lowercase
and uppercase letters denote vectors and matrices, respectively. The operators $\tr\{\cdot\}$, vec$(\cdot)$,
$(\cdot)^\mathrm{T}, (\cdot)^*,$ $(\cdot)^{\rm H}$, $(\cdot)^{-1}$, $\Re(\cdot)$, and $\Im(\cdot)$ stand for the trace, the vectorization, the transpose,
the conjugate, the Hermitian, the inverse, the real part, and the imaginary part of their arguments, respectively. For any vector $\xb$, $\text{diag}\{\xb\}$ returns a diagonal square matrix whose diagonal consists of the elements of $\xb$. For any matrix $\Xb$, $[\Xb]_{\text{diag}}$ represents
a diagonal matrix with the same diagonal elements of $\Xb$. The operation $\otimes$ denotes the
Kronecker product. $\Ib_{N}$ denotes an identity matrix of size $N \times N$. 
\section{Background of Bayesian Optimization}\label{preli}
Consider the following problem of finding a global minimizer (or maximizer) of an unknown objective function
\begin{equation}\label{eq:bo}
  \xb^* = \arg \mathop{\min_{\xb \in \mathcal{X}}} f(\xb),
\end{equation}
where $\mathcal X$ is the compact set of the domain, i.e., a hyper-rectangle $\mathcal X  =\{\xb\in\mathbb{R}^D: a_i\leq x_i \leq b_i\}$. We can only  query at some $\xb_i \in \mathcal{X}$ and obtain the corresponding observation $f(\xb_i)$. Note that $f$ may be nonconvex and the gradient information is not available.
Since the objective function is unknown, the Bayesian strategy is to treat it as a random function and place a prior over it. The prior captures beliefs about the behavior of the function. After gathering the function evaluations, which are treated as data, the prior is updated to form the posterior distribution of the objective function, which in turn, is used to construct an acquisition function that determines the next query point.

In particular, the widely used Gaussian prior is adopted, i.e., $f(\xb_{1:t}) = [f(\xb_1),f(\xb_2),\dots, f(\xb_t)]^{\rm T} \sim\mathcal{N}(\zero,\Kb_t)$, where $\Kb_t$ is the $t \times t $ kernel matrix with $\Kb_t(i,j) = k( \|\xb_i- \xb_j\|), i,j=1, ..., t. $
Two popular kernels are the squared exponential (SE) kernel and the M\`{a}tern kernel, given respectively by
\begin{align}
 k_{\text{SE}}(a) &= \exp\left(\frac{-a^2}{2h^2}\right), \\
 k_{\text{M\`{a}tern}}(a) &= \frac{2^{1-\nu}}{\Gamma(\nu)}\left(\frac{\sqrt{2\nu}a}{h}\right)^\nu B_{\nu}\left(\frac{\sqrt{2\nu}a}{h}\right),
\end{align}
where $\Gamma(\cdot)$ and $B_\nu(\cdot)$ are the Gamma function and the $\nu$-th order Bessel function, respectively; $h$ is a hyper-parameter.

In Bayesian optimization, at the $t$-th iteration, we have samples $\mathcal{D}_{1:t} = \{(\xb_{i},f(\xb_{i}))\}_{i=1}^t$ and we are interested in inferring the value of $f(\xb_{t+1})$ at the next query point $\xb_{t+1}$.
By the Gaussian prior assumption,
 \begin{equation}\label{joint_Gau}
   \begin{bmatrix} f(\xb_{1:t}) \\ f(\xb_{t+1}) \end{bmatrix} \sim \mathcal{N} \left(\zero, \begin{bmatrix} \Kb_t & \kb_{t+1} \\ \kb^{\rm T}_{t+1} & k(0)\end{bmatrix}\right)
 \end{equation}
 with $\kb_{t+1} = [k(\|\xb_{t+1}-\xb_1\|),k(\|\xb_{t+1}-\xb_2\|),\dots, k(\|\xb_{t+1}-\xb_t\|)]^{\rm T}$, and the property of Gaussian distribution, we can write
 \begin{equation}\label{x_n+1}
   f(\xb_{t+1})|\mathcal{D}_{1:t} \sim \mathcal{N}(\mu_t(\xb_{t+1}), \sigma^2_{t}(\xb_{t+1})),
 \end{equation}
 where
 \begin{align}
 \mu_t(\xb_{t+1}) &=  \kb^{\rm T}_{t+1}\Kb_t^{-1} f(\xb_{1:t}), \label{eq:mu1}\\
 \sigma^2_t(\xb_{t+1}) &= k(0) - \kb_{t+1}^{\rm T}\Kb_t^{-1}\kb_{t+1}. \label{eq:sigma1}
 \end{align}
Note that the computation complexity of \eqref{eq:mu1} and \eqref{eq:sigma1} is high since the dimensions of the matrix and vectors grow with $t$. In order to reduce the complexity, we can set a window size $W$ and compute $\mu_t(\xb
_{t+1})$ and $\sigma^2_t(\xb_{t+1})$ based on $\{(\xb_{i},f(\xb_{i}))\}_{i=t-W+1}^t$ instead of $\{(\xb_{i},f(\xb_{i}))\}_{i=1}^t$. Define the $W \times 1$ vector ${\bar \kb}_{t+1}  = [k(\|\xb_{t+1}-\xb_{t-W+1}\|),k(\|\xb_{t+1}-\xb_{t-W+2}\|),\dots, k(\|\xb_{t+1}-\xb_t\|)]^{\rm T}$ and the $W \times W$ kernel matrix ${\bar \Kb} _t$ such that
$\bar \Kb_t (i, j) = k( \|\xb_{t-W+i}- \xb_{t-W+j}\|), i,j=1, \dots, W.$   Then we can replace \eqref{eq:mu1}-\eqref{eq:sigma1} by
 \begin{align}
 \mu_t(\xb_{t+1}) &=  \bar\kb_{t+1}^{\rm T}\bar\Kb^{-1}_{t} f(\xb_{t-W+1:t}), \label{eq:mu}\\
 \sigma^2_t(\xb_{t+1}) &= k(0) - \bar\kb_{t+1}^{\rm T}\bar\Kb^{-1}_{t}\bar\kb_{t+1}. \label{eq:sigma}
 \end{align}
Note that in \eqref{eq:mu}-\eqref{eq:sigma} the matrix $\bar\Kb_t$ and vector  $\bar\kb_{t+1}$ are both functions of $\xb_{t+1}$.
Then, the posterior mean and variance  in \eqref{eq:mu}-\eqref{eq:sigma} are used to construct the following acquisition function
%

\begin{algorithm}[t]
\caption{The Bayesian optimization algorithm for solving \eqref{eq:bo}}
\label{alg:A}
\begin{algorithmic}[1]
\For{ $t= 1, 2, \dots, T$}
    \State Calculate $\mu_{t+1}$ and $\sigma_{t+1}$ using \eqref{eq:mu} and \eqref{eq:sigma};
    \State Find $\xb_{t+1} = \arg \mathop{\min_{\xb \in \mathcal{X}}} \varphi_{t+1}(\xb|\mathcal{D}_{1:t})$ by performing the grid search algorithm;
    \State Evaluate $ f(\xb_{t+1})$;
\EndFor \\
\algorithmicensure ~ $\xb^* = \xb_T$.
\end{algorithmic}
\end{algorithm}
\begin{equation}\label{GP-LCB}
  \varphi_{t+1}(\xb) = \mu_{t}(\xb) - \sqrt{\beta_{t+1}}\sigma_{t}(\xb),
\end{equation}
where $\beta_{t+1}$ is a hyper-parameter. The next query point $\xb_{t+1}$ is then given by
\begin{equation}\label{max_ac}
  \xb_{t+1} = \arg \mathop{\min_{\xb\in \mathcal{X}}} \varphi_{t+1}(\xb).
\end{equation}
In practice, \eqref{GP-LCB} can be solved using various numerical methods, e.g., the grid search algorithm in \cite{DIRECT}. The Bayesian optimization procedure for solving \eqref{eq:bo} is summarized in Algorithm $1$. 

\section{System Description and Problem Formulations}\label{Sys}
In this section, we first introduce the system model, and then provide problem formulations for RIS-assisted downlink data transmission and power transfer, respectively.
\subsection{System Model}
\begin{figure}[t]
  \centering
  \includegraphics[width=3.5in]{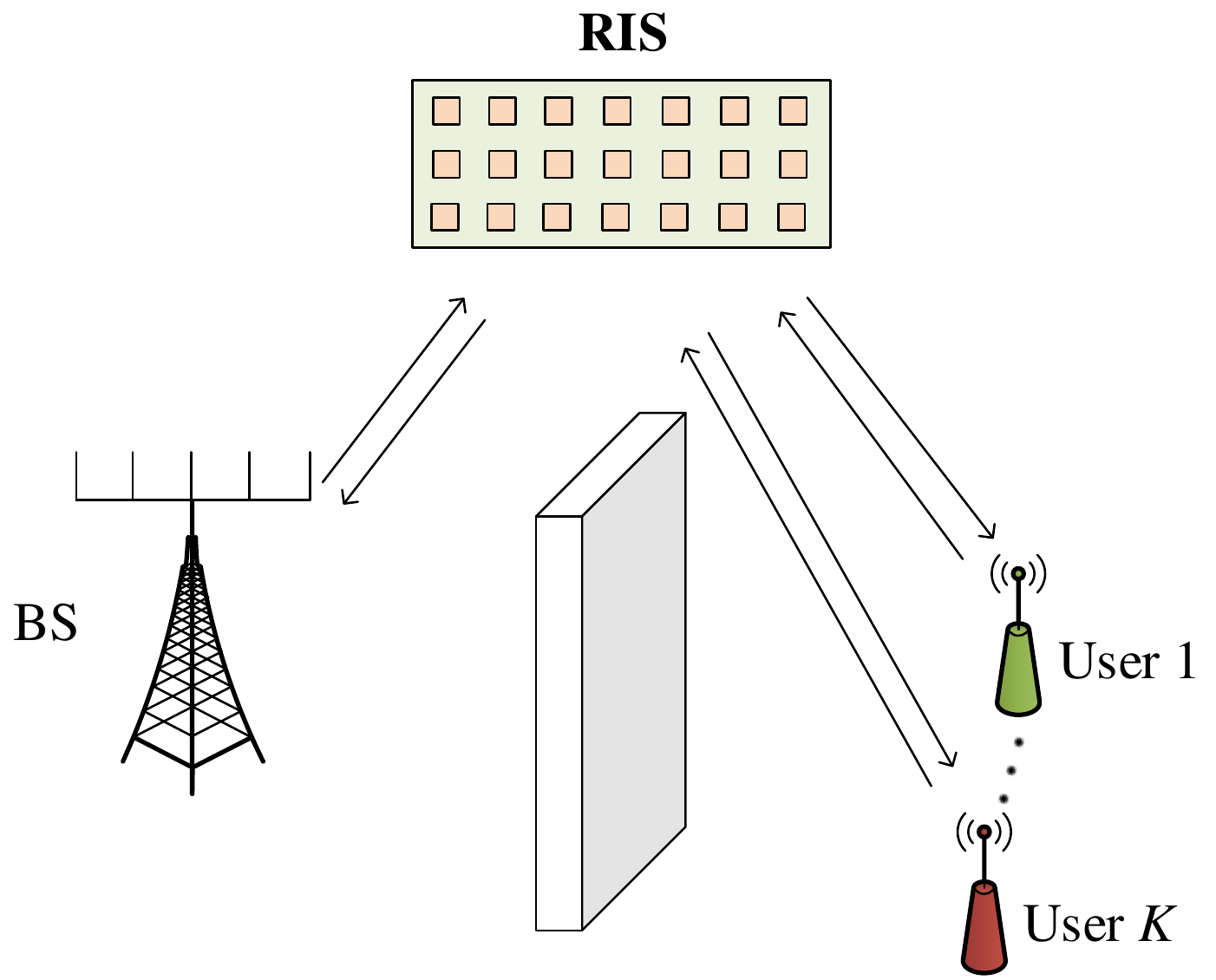} \\
  \caption{The RIS-assisted multiuser MISO system.}\label{miso_multiple_user}
\end{figure}

The system model is shown in Fig. \ref{miso_multiple_user}. We consider an RIS-assisted multi-user MISO system, where the BS equipped with $M$ antennas serves $K$ single-antenna users with the help of an $N$-element RIS. There is no direct link between the BS and users. The RIS is usually installed on a surrounding wall to assist the transmissions between the BS and the users. Each element of the RIS is configurable and programmable via an RIS controller. Moreover, the RIS is connected to a smart controller, which can be used to exchange the information with the BS and the users. The transmitted signal at the BS is given by
\begin{equation}\label{bs_trans}
  \xb = \Wb\ssb,
\end{equation}
where $\Wb = [\wb_1,\wb_2,\dots,\wb_M]^{\rm T} \in \mathbb{C}^{M\times K}$ is the linear precoding matrix at the BS; $\ssb = [s_1, s_2, \dots, s_K]^{\rm T}$ contains independent and identically distributed (i.i.d.) transmitted data symbols to users each with zero mean and unit variance. The transmission power constraint is given by  $\tr\{\Wb^{\rm H}\Wb\}\leq P$. Then, the received signal at the RIS is written as
\begin{equation}\label{ris_re}
  \yb = \Hb\xb,
\end{equation}
where $\Hb\in \mathbb{C}^{N\times M}$ is the channel matrix from the BS to the RIS. The RIS is a reflection device and it effectively applies a phase shift to each element of the received signal $\yb$ and then forwards it to the users. The received signal at users is then given by
\begin{equation}\label{user_re}
  \rb = [r_1,r_2,\dots,r_K]^{\rm T} = \Fb\mPhi\Hb\Wb\ssb + \ub,
\end{equation}
where $\Fb = [\fb_1,\fb_2,\dots,\fb_K]^{\rm T}\in \mathbb{C}^{K\times N}$ denotes the channel matrix between the RIS and the users; $\mPhi = \text{diag} \{\mphi\}$ with $\mphi=[\phi_1, \phi_2, \dots, \phi_N]^{\rm T}$ and $\phi_n = e^{j\theta_n}$ is the phase shift at the $n$-th RIS antenna element, $n= 1,2,\dots, N$;  $\ub = [u_1,u_2,\dots,u_K]^{\rm T}$ contains i.i.d. zero-mean circularly symmetric complex Gaussian noise at the user, i.e., $u_k \sim \mathcal{CN}(0, \gamma^2)$.
Finally, each user $k$ scales its received signal $r_k$ by $c_k \in \Cb$, which is represented by
\begin{equation}\label{user_fit}
  \hat\ssb = \Cb\rb = \Cb\Fb\mPhi\Hb\Wb\ssb + \Cb\ub,
\end{equation}
where $\Cb = \text{diag}\{c_1,c_2,\dots,c_K\}$.
\subsection{Problem Formulations}
The above RIS-assisted system can be employed for downlink data transmission or power transfer. For either case, we need to design the matrices $\{ \Wb, \mPhi, \Cb \}$, which depend on the channels $\{ \Hb, \Fb \}$.  The key challenge is then to obtain the optimal design without knowing the channels; and for that we resort to the tool of Bayesian optimization. In what follows, we first formulate the optimization problems for data transmission and power transfer.

\subsubsection{Downlink Data Transmission}

For downlink data transmission, one objective is to minimize the sum MSE,  given by
\begin{align}\label{mse}
  f_{\text{MSE}}(\Wb,\mPhi,\Cb) &\triangleq\mathbb{E}\{\|\hat\ssb - \ssb\|^2\} \\
  &= \mathbb{E}\{\|(\Cb\Fb\mPhi\Hb\Wb-\Ib) \ssb + \Cb\ub\|^2\} \\
               &= \tr\{(\Cb\Fb\mPhi\Hb\Wb - \Ib)^{\rm H}(\Cb\Fb\mPhi\Hb\Wb - \Ib)\} + \gamma^2\tr\{\Cb^{\rm H}\Cb\} \\
               &= \tr\{\Wb^{\rm H}\Hb^{\rm H}\mPhi^{\rm H}\Fb^{\rm H}\Cb^{\rm H}\Cb\Fb\mPhi\Hb\Wb+\Ib\}- \tr\{\Wb^{\rm H}\Hb^{\rm H}\mPhi^{\rm H}\Fb^{\rm H}\Cb^{\rm H}\} \notag\\
               &\qquad -\tr\{\Cb\Fb\mPhi\Hb\Wb\} + \gamma^2\tr\{\Cb^{\rm H}\Cb\}.
\end{align}
Then the optimization problem becomes
\begin{subequations}\label{opt_1}
\begin{align}
\qquad	 \mathop{\min}_{\{\Wb, \mPhi, \Cb\}} ~~& f_{\text{MSE}}(\Wb,\mPhi,\Cb)  \label{eq:maxuplink_multi}\\
	\st  ~~&  \tr\{\Wb^{\rm H}\Wb\} \leq P \label{const_pow}\\
         ~~&  0 \leq \theta_n \leq 2\pi, ~~\forall n= 1,2, \dots, N \label{const_phase}.
\end{align}
\end{subequations}
Note that since the channel state information (CSI) $\{ \Hb, \Fb\}$ is not available, the above optimization problem cannot be solved analytically. 

\remark{To evaluate the expectation in \eqref{eq:maxuplink_multi},  at each iteration $t$,  we can transmit $\kappa$ pilot symbol vectors $\ssb(1),\dots,\ssb(\kappa)$, and evaluate an approximate objective value
\begin{equation}
f_{\text{MSE}}( \Wb, \mPhi, \Cb) \approx \frac 1 \kappa \sum_{n=1}^\kappa \| \hat \ssb (n) - \ssb(n) \|^2. \label{eq:est_mse}
\end{equation}
Simulation results in Sec.V indicate that one pilot symbol vector, e.g., $\kappa=1$ suffices. }


To take into account the fairness among users, we can also minimize the maximum of individual MSE. In particular, the MSE of user $k$ is given by
\begin{align}
 f_{\text{MSE}_k}(\wb_k,\mPhi,c_k) &\triangleq\mathbb{E}\{|\hat s_k - s_k|^2\} \\
  &= \mathbb{E}\{|(c_k \fb_k \mPhi\Hb\sum_{k=1}^{M} \wb_k s_k + c_k u_k - s_k |^2\} \\
               &= (c_k\fb_k\mPhi\Hb\wb_k - 1)^{\rm H}(c_k\fb_k\mPhi\Hb\wb_k - 1) + \gamma^2 c_k^{\rm H}c_k\\
               &= \wb_k^{\rm H}\Hb^{\rm H}\mPhi^{\rm H}\fb_k^{\rm H}c_k^{\rm H}c_k\fb_k\mPhi\Hb\wb_k- 2\Re\{\wb_k^{\rm H}\Hb^{\rm H}\mPhi^{\rm H}\fb_k^{\rm H}c_k^{\rm H}\}  + \gamma^2 c_k^{\rm H}c_k+1 .
\end{align}
Then we can consider the following min-max formulation:
\begin{subequations}\label{opt_1_1}
\begin{align}
\qquad	 \mathop{\min}_{\{\wb_k, \mPhi, c_k\}} ~~\mathop{\max}_{k} ~~ & f_{\text{MSE}_k}(\wb_k,\mPhi,c_k)  \label{eq:maxuplink_multi}\\
	\st  ~~&  \eqref{const_pow}, \eqref{const_phase}. \label{eq:re_const}
\end{align}
\end{subequations}
Note that the objective function in \eqref{eq:maxuplink_multi} is not smooth which is hard to be optimized via Bayesian optimization. Therefore we employ the following smooth approximation to the maximization operator \cite{Smooth}:
\begin{align}\label{eq:maxf}
  \max \{x_1,x_2,\dots, x_K\} \approx \eta\ln \sum_{k=1}^{K} \exp(\frac{f_{\text{MSE}_k}(\wb_k,\mPhi,c_k)}{\eta}),
\end{align}
where $\eta>0$ is a parameter. Large $\eta$ leads to high accuracy of the approximation, but it
also causes the problem to be nearly ill-conditioned. When $\eta$ is chosen appropriately,  we can use the following smooth reformulation of \eqref{eq:maxuplink_multi}-\eqref{eq:re_const}:
\begin{subequations}\label{opt_1_2}
\begin{align}
  \qquad	 \mathop{\min}_{\{\wb_k, \mPhi, c_k\}} ~~&\eta\ln \sum_{k=1}^{K} \exp(\frac{f_{\text{MSE}_k}(\wb_k,\mPhi,c_k)}{\eta})  \label{eq:maxuplink_multi_re}\\
	\st  ~~&  \eqref{const_pow}, \eqref{const_phase}. \label{eq:re_const_1}
\end{align}
\end{subequations}
\subsubsection{Downlink Power Transfer}

For downlink wireless power transfer, the receiving filter $\Cb$ is not needed at the users. The total harvested power at the users is defined as
\begin{align}
f_{\text{p}}(\Wb, \mPhi) &\triangleq \mathbb{E}\{\rb\rb^{\rm H}\} \\
 & = \tr\{\Fb\mPhi\Hb\Wb\Wb^{\rm H}\Hb^{\rm H}\mPhi^{\rm H}\Fb^{\rm H}\} +  \gamma^2 K.
\end{align}
The optimization problem is then
\begin{subequations}\label{opt_2}
\begin{align}
\qquad	 \mathop{\max}_{\{\Wb, \mPhi\}} ~~& f_{\text{p}}(\Wb, \mPhi)   \label{eq:max_obj_2}\\
	\st  ~~&  \eqref{const_pow}, \eqref{const_phase}.
\end{align}
\end{subequations}

We can also  maximize the minimum of individual harvested power. In particular, the harvested power at user $k$ is defined as
\begin{align}
f_{\text{p}_k}(\Wb, \mPhi) &\triangleq \mathbb{E}\{r_k r_k^{\rm H}\} \\
 & = \fb_k\mPhi\Hb\Wb\Wb^{\rm H}\Hb^{\rm H}\mPhi^{\rm H}\fb_k^{\rm H} +  \gamma^2.
\end{align}
Then we consider the following max-min formulation:
\begin{subequations}\label{opt_2_1}
\begin{align}
\qquad	 \mathop{\max}_{\{\Wb, \mPhi\}} ~~\mathop{\min}_{k} ~~& f_{\text{p}_k}(\Wb, \mPhi)   \label{eq:max_obj_2}\\
	\st  ~~&  \eqref{const_pow}, \eqref{const_phase}.
\end{align}
\end{subequations}
Similarly as before, a smooth reformulation of (34a)-(34b) is given by
\begin{subequations}\label{opt_2_2}
\begin{align}
  \qquad	 \mathop{\max}_{\{\Wb, \mPhi\}} ~~&\eta\ln \sum_{k=1}^{K} \exp(\frac{f_{\text{p}_k}(\Wb,\mPhi)}{\eta}) \\
	\st  ~~&  \eqref{const_pow}, \eqref{const_phase}.
\end{align}
\end{subequations}

\section{Downlink Transmission Designs With Unknown CSI}\label{BO_app}
In this section, we propose a  Bayesian optimization based approach to solve the total MSE minimization problem in \eqref{opt_1} based on observing the sample MSE. Note that the same approach can be used to solve the other problems formulated in Sec. \ref{Sys}.
\subsection{Spherical  Representation of Variables}
We treat the objective in \eqref{opt_1} as a black-box function with input variables  $\{\Wb$, $\mPhi$, $\Cb\}$. In order to employ Bayesian optimization, the input variables should be real-valued and the optimization should be unconstrained. We first transform  $\{\Wb$, $\mPhi$, $\Cb\}$ into real-valued vectors. The phase shift $\mPhi$ consists of $N$ real variables $\mtheta = [\theta_1,\theta_2,\dots,\theta_N]^{\rm T} \in [0, 2\pi]^{N}$.

For the  beamforming matrix $\Wb \in \mathbb{C}^{M\times K}$, we can first vectorize it as
$\wb = \vec (\Wb) \in \mathbb{C}^{MK}$ and then convert it to real-valued as
$\tilde\wb = [\Re\{\wb^{\rm T}\}, \Im\{\wb^{\rm T}\}]^{\rm T}\in \mathbb{R}= [ \tilde w_1, \tilde w_2,\ldots, \tilde w_{2MK} ] ^{2MK}$. Then the power constraint  \eqref{const_pow} becomes
\begin{equation}\label{pow_expand}
  \tilde w_{1}^2 + \tilde w_{2}^2 + \dots + \tilde w_{2MK}^2 \leq P.
\end{equation}
In order to remove this constraint, we parameterize the vector $\tilde\wb$ by spherical coordinates $\mpsi = [\psi_1,\psi_2,\dots,\psi_{2MK-1}]^{\rm T} \in [0,\pi]^{2MK-1}$ as follows
\begin{align}
\tilde w_m &= \sqrt{P}\cos\psi_m\prod_{n=1}^{m-1}\sin\psi_n, \quad m= 1,2,\dots,2MK-1, \label{eq:e}\\
\tilde w_{2MK} & = \sqrt{P}\sin\psi_{2MK-1}\prod_{n=1}^{2MK-2}\sin\psi_n, \quad m= 2MK. \label{eq:rf}
\end{align}
It can be easily checked that $\tilde \wb$ defined by
\eqref{eq:e}-\eqref{eq:rf} always satisfies \eqref{pow_expand}.
Moreover, the filtering matrix $\Cb$ consists of $K$ complex variables $\cb = [c_1,c_2,\dots,c_K]^{\rm T}\in \mathbb{C}^{K}$, which can be represented as $2K$ real variables  $\tilde\cb=[ \Re\{\cb^{\rm T}\},\Im\{\cb^{\rm T}\}]^{\rm T}\in \mathbb{R}^{2K}$. Note that we observe from simulations that its elements $\tilde c_k\in [-1,1]$ under the Rayleigh fading channel model. Therefore, we can also represent $\tilde\cb$ using the spherical coordinate, i.e., $\tilde c_k = \cos \gamma_k$, with $\gamma_k \in [0,\pi].$
Hence the constrained optimization  problem in \eqref{opt_1} with complex-valued variables is converted to an unconstrained optimization with $D = 2(M+1)K+N-1$ real-valued variables
\begin{equation}
\qquad	 \mathop{\min}_{\xb} ~~ f_{\text{MSE}}(\xb),
\end{equation}
with $\xb = [\mtheta^{\rm T}, \mpsi^{\rm T},\boldsymbol\gamma^{\rm T}]^{\rm T}$.

\subsection{Optimization via Coordinate Decomposition}
When the dimension $D$ is high, the optimization in \eqref{max_ac} is hard to solve.  A typical approach is to partition the $D$-dimensional decision variable $\xb$ into $L$ non-overlapping  segments: $\xb^{(1)}, ..., \xb^{(L)}$. 
We postulate that the objective function can be decomposed as the following:
\begin{equation}\label{eq:decom}
f(\xb) \approx  f^{(1)}(\xb^{(1)}) + f^{(2)}(\xb^{(2)})+ \dots +f^{(L)}(\xb^{(L)}).
\end{equation}
Moreover, given $\{(\xb_{i},f(\xb_{i}))\}_{i=t-W+1}^t$, we further assume that $f^{(\ell)}(\xb^{(\ell)}_{t+1})$ and $f(\xb_{t-W+1:t})$ are jointly Gaussian. Similar to \eqref{eq:mu}-\eqref{eq:sigma}, we obtain the posterior mean and the posterior variance of $f^{(\ell)}(\xb^{(\ell)}_{t+1})$ as
 \begin{align}
 \mu_t^{(\ell)}(\xb^{(\ell)}_{t+1}) &=  (\bar\kb_{t+1}^{(\ell)})^{\rm T}\bar\Kb^{-1}_{t} f(\xb_{t-W+1:t}), \label{eq:mu_1}\\
 (\sigma^{(\ell)}_{t}(\xb_{t+1}^{(\ell)}))^2 &= k(0) - (\bar\kb^{(\ell)}_{t+1})^{\rm T}\bar\Kb^{-1}_{t}\bar\kb^{(\ell)}_{t+1}, \label{eq:sigma_1}
 \end{align}
where $\bar\kb_{t+1}^{(\ell)} = [k(\|\xb^{(\ell)}_{t+1}-\xb^{(\ell)}_{t-W+1}\|),k(\|\xb^{(\ell)}_{t+1}-\xb^{(\ell)}_{t-W+2}\|),\dots, k(\|\xb^{(\ell)}_{t+1}-\xb^{(\ell)}_t\|)]^{\rm T}$. The acquisition function $\varphi^{(\ell)}_{t+1}$ of the segment $\xb^{(\ell)}$ is then given by
\begin{align}\label{eq:acq_add}
  \varphi^{(\ell)}_{t+1}(\xb^{(\ell)}) &= \mu^{(\ell)}_{t}(\xb^{(\ell)}) -\sqrt{\beta_{t+1}}\sigma_{t}^{(\ell)}(\xb^{(\ell)}).
 \end{align}
And the next query segment $\xb_{t+1}^{(\ell)}$ is given by
\begin{equation}\label{eq:seg_1}
  \xb^{(\ell)}_{t+1} = \arg\mathop{\min}_{\xb^{(\ell)}}\varphi^{(\ell)}_{t+1}(\xb^{(\ell)}), \quad \ell = 1,\dots, L.
\end{equation}
Finally we combine all segments to obtain the query point $\xb_{t+1}$ and  the corresponding objective function value $f(\xb_{t+1})$. 
\begin{figure}[t]
  \centering
  \includegraphics[width=6in]{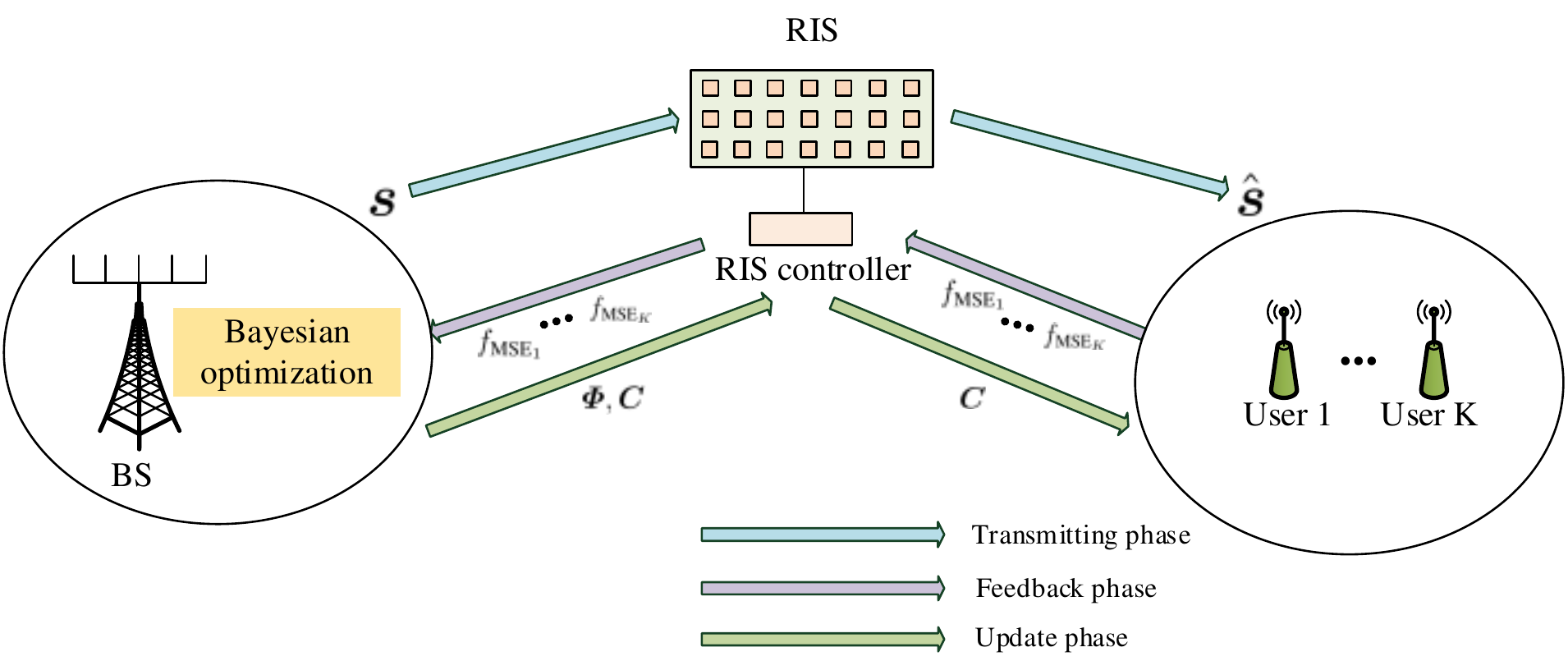} \\
  \caption{The proposed scheme via Bayesian optimization in the downlink data transmission.}\label{bo_model}
\end{figure}

Moreover, we can further consider multiple, say $Q$, random partitions of $\xb$. For each partition, we compute the query point $\xb_{t+1}$ according to the above procedure  and the objective value $f(\xb_{t+1})$. Then we choose the partition that results in the minimum value of objective value. 
\begin{algorithm}[t]
\caption{The proposed training scheme based on Bayesian optimization}
\label{alg:BO_add}
\begin{algorithmic}[1]
\State {\textbf{Initialization process:}}
\For{ $t = -W+1,  \dots, 0$}
\State The BS, the RIS, and the users randomly initialize $\xb_t$.
\State The BS transmits pilot $\ssb(t)$ to users.
\State The users estimate the objective value in \eqref{eq:est_mse} and feed it back to the BS.
\EndFor
\State $t=1$
\For { each one of the $Q$ possible partitions of $\xb$}
\State The BS computes $\xb_1$ in \eqref{eq:seg_1}.
\State The BS transmits pilot $\ssb(1)$ to users.
\State The users estimate the objective value $f(\xb_1)$ and feed it back to the BS.
\EndFor
\State The BS chooses the partition corresponds to minimum of $f(\xb_1)$.
\State \textbf{Updating process:}
\For{ $t= 2, \dots, T$}
\State The BS transmits pilot $\ssb(t)$ to users.
\State The users estimate the objective value $f(\xb_t)$ and feed it back to the BS.
     \For {$\ell = 1,2,\dots,L$}
    \State Compute $\mu_{t}^{(\ell)}$ and $\sigma_{t}^{(\ell)}$ using  \eqref{eq:mu_1} and \eqref{eq:sigma_1}, respectively;
    \State Update $\xb_{t+1}^{(\ell)}$ in \eqref{eq:seg_1} using the grid search algorithm;


    \EndFor
    \State $\xb_{t+1} = \cup_{\ell=1}^{L}\xb_{t+1}^{(\ell)}$;
    \State The BS transmits the updated $\mPhi$ and $\Cb$ to the RIS and the users;

\EndFor\\
\algorithmicensure ~ $\{\Wb,\mPhi,\Cb\}$.
\end{algorithmic}
\end{algorithm}
In summary, the practical implementation of the proposed transmission scheme is shown in Fig. \ref{bo_model}. The total training procedure consists of two stages: the initial stage and the updating stage.
First, in the initialization process, the BS, the RIS, and the users randomly initialize input vector $\xb_t$  for $W$ times. After each initialization, the BS transmits the pilot symbol $\ssb(t)$ to the users. Then the users estimate the objective value in \eqref{eq:est_mse} and feed it back to the BS. Based on the $W$ initial sample points and the corresponding objective value, we randomly generate $Q$ possible partitions of $\xb$. Then the BS computes  $\xb_1$ in \eqref{eq:seg_1} and transmits pilot $\ssb(1)$. The users estimate the objective value $f(\xb_1)$ and feed it back to the BS. Then the BS chooses the partition corresponds to a minimum of $f(\xb_1)$.

%

In the updating process, the BS transmits the pilot symbol $\ssb(t)$ to the users. Then the users estimate the objective value $f(\xb_t)$ and feedback to the RIS and the BS. The main difference with the initial stage is that the Bayesian optimization is operated  at each time to obtain the next query point $\xb_{t+1}$. Then the BS informs the RIS and the users of the updated $\mPhi$ and $\Cb$, respectively. Finally, by repeating this updating process, the sum MSE is minimized. The proposed Bayesian optimization based training scheme is shown in Algorithm $2$.
\begin{figure}[t]
  \centering
  \includegraphics[width=5in]{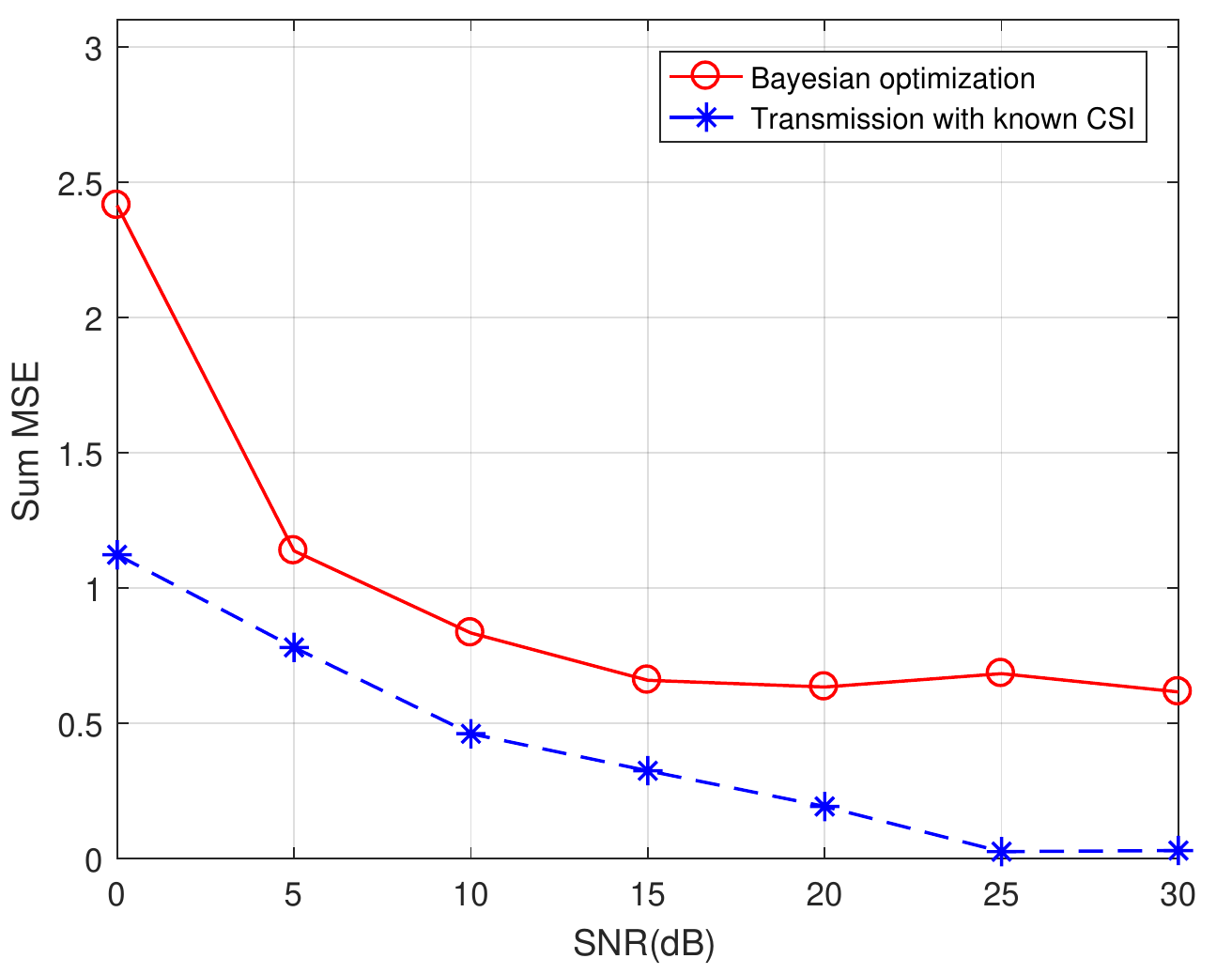} \\
  \caption{The sum MSE performance of the proposed Bayesian optimization method and the known CSI scheme is evaluated in the downlink data transmission. The proposed ``Bayesian optimization'' has an acceptable performance with the ``Transmission with known CSI'' in the Appendix. }\label{miso_multiple_mse}
\end{figure}
\begin{figure}[t]
  \centering
  \includegraphics[width=5in]{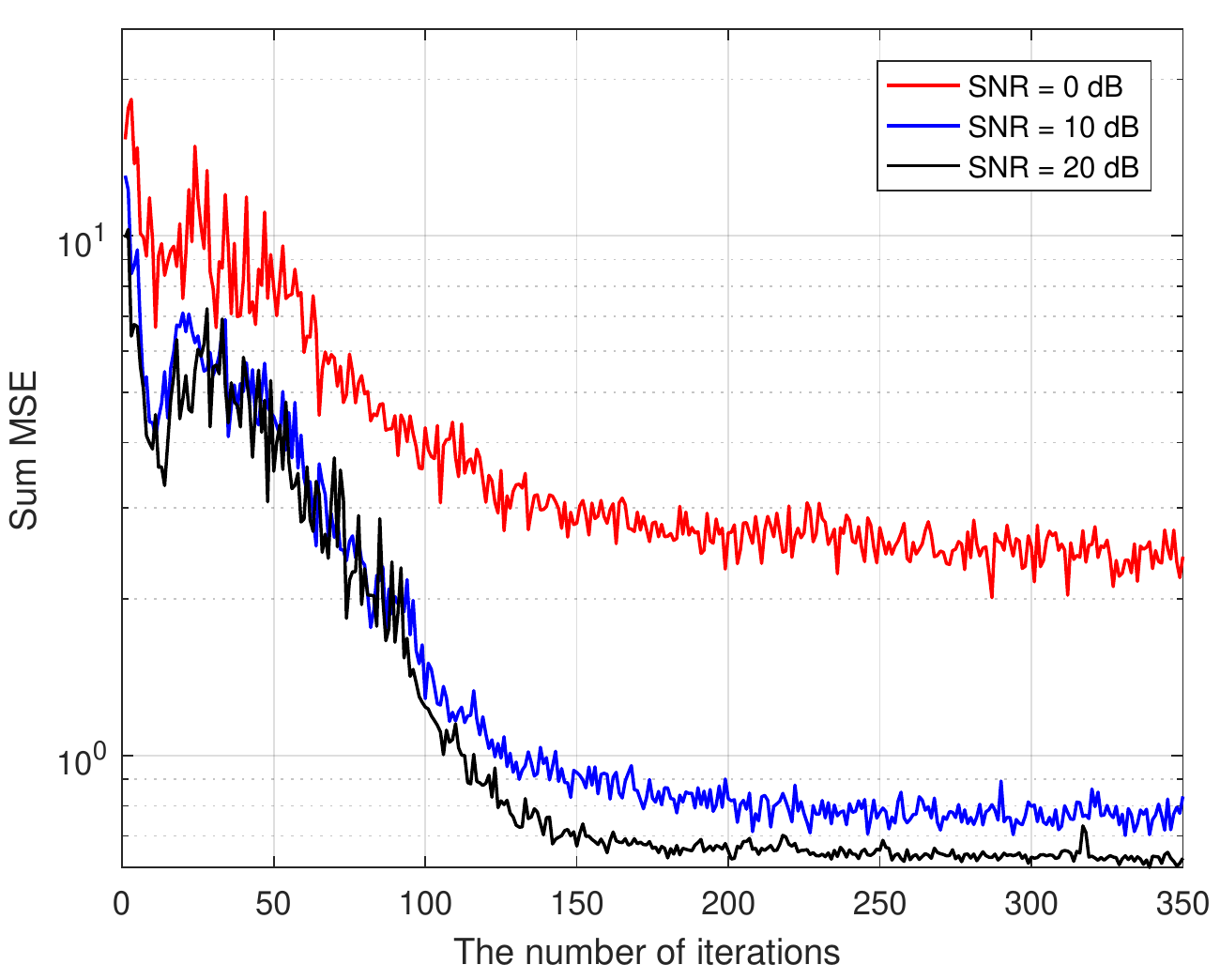} \\
  \caption{The convergence time for the proposed transmission scheme is evaluated in the downlink data transmission. It shows that the proposed scheme can achieve convergence through iterations.}\label{mimo_multiple_iter}
\end{figure}
\section{Simulation Results}\label{simula}
In this section, we present numerical results to verify the performance of the proposed transmission scheme based on Bayesian optimization.  The BS is equipped with $M = 2$ antennas. The number of elements in RIS is $N =2$  if there is no specific introduction. The parameter $\beta_{t+1}$ in \eqref{GP-LCB} is set to $0.4\log(2t+2)$. The window size is $W = 20$.  The SE kernel is chosen in Bayesian optimization and the hyper-parameter $h$ is learned by the maximum likelihood estimation in \cite{brochu2010tutorial}. The number of partition $Q$ is equal to the input dimension $D$. The smooth parameter is $\eta = 50$. The simulation results are obtained by taking an average over 1000 random realizations and the maximum iteration time $T = 350$.

Furthermore, for the initialization of the proposed algorithm, we set the initial phase shift of each reflecting element follows a uniform distribution, i.e., $\theta_i \in \mathcal{U}(0,2\pi)$. Similarly,  the filtering element is $ \gamma_k \in \mathcal{U}(0,\pi)$. For the initialization of $\mpsi$, we first randomly generate the vector $\tilde \wb$ satisfying the power constraint \eqref{pow_expand}, and then compute the corresponding $\mpsi$ by inverting the spherical coordinate transformation \eqref{eq:e}, \eqref{eq:rf} as follows:
\begin{align}
\psi_1 &= \arccos\left(\frac{\tilde w_1}{\sqrt{P}}\right), \\
\psi_{m} &= \arccos\left(\frac{\tilde w_m}{\sqrt{P}\Pi_{n=1}^{m-1}\sin\psi_n}\right), \quad \text{for}~ m = 2,\dots,2MK-1.
\end{align}
\begin{figure}[t]
  \centering
  \includegraphics[width=5in]{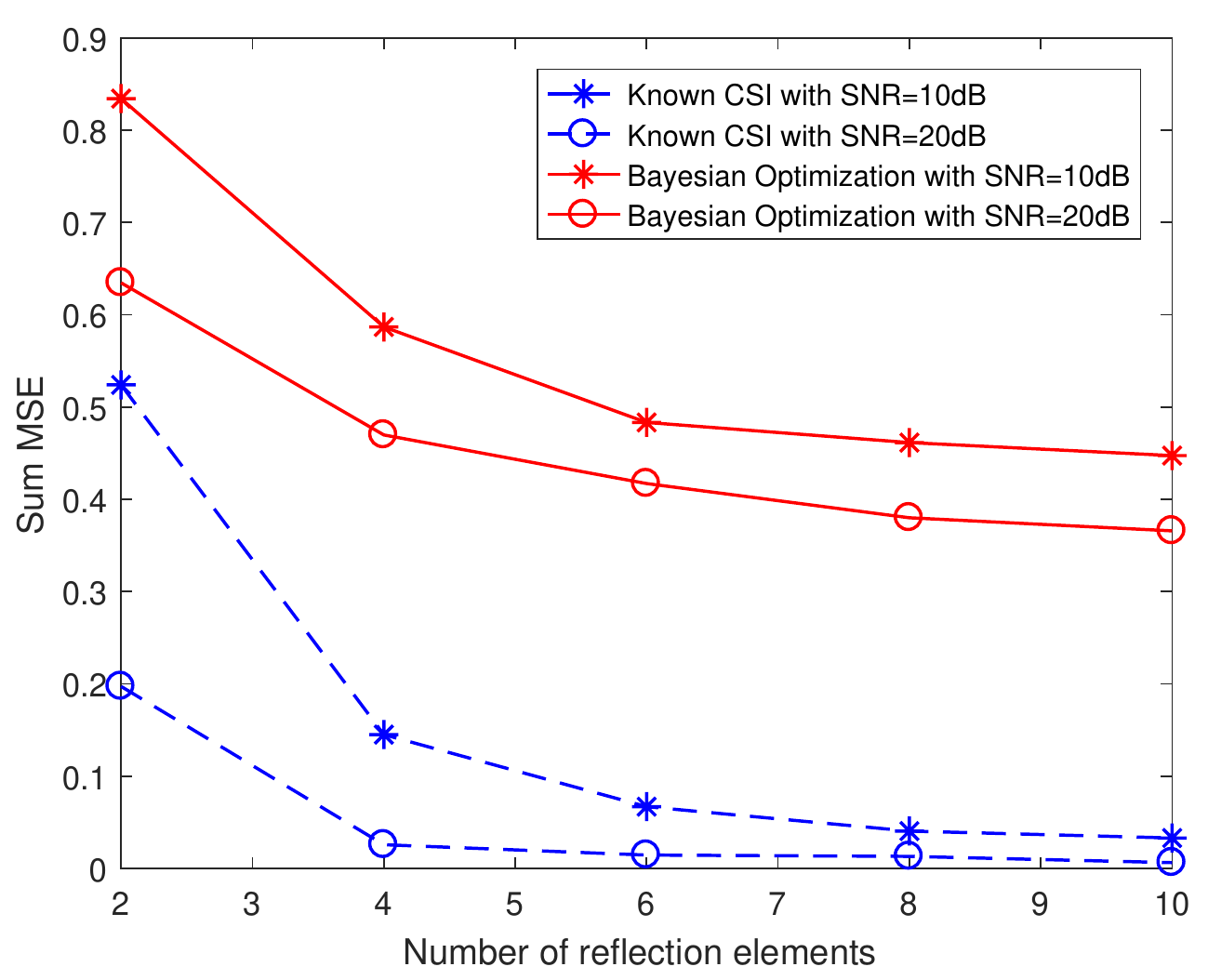} \\
  \caption{The sum MSE performance of different reflection elements for the proposed scheme and the transmission scheme with known CSI is evaluated in the downlink data transmission. When $\mathsf{SNR} = 20$ dB, the sum MSE of the proposed ``Bayesian optimization'' with ten reflecting elements  is nearly same as the ``Known CSI'' with two reflecting elements in Appendix.}\label{n_element}
\end{figure}
\subsection{Downlink Data Transmission}
In the downlink data transmission, the Rayleigh fading is assumed for each link.  The sum MSE performance of the proposed Bayesian optimization scheme and the known CSI scheme is shown in Fig. \ref{miso_multiple_mse}. The derivation of the benchmark ``Transmission with known CSI'' is shown in Appendix. As we can see, the proposed Bayesian optimization scheme is acceptable compared with the scheme with known CSI, especially at the medium SNR ($5<\mathsf{SNR}<15$). It is worth noting that the sum MSE performance degrades in the low SNR regime since the noise varies dramatically in each feedback. Moreover, the sum MSE is convergent at 0.7 and does not reduce any more. This is because the objective function is approximated as the sum of low-dimension subfunctions, which makes the performance loss.

The convergence time for the proposed scheme is evaluated in Fig. $\ref{mimo_multiple_iter}$. The SNR is set to $\{0 ~\text{dB},~10 ~\text{dB},~20 ~\text{dB}\} $. Although the sum MSE is high at the initial samples, the sum MSE gradually convergent after 150 times in both the high or the low SNR regimes. This shows that the number of iterations of the proposed algorithm is acceptable in practice. Note that the more iterations guarantee better performance since the objective value of the new sample point may be stable at the local optimal. Moreover, it can be seen that slight fluctuations appear at the convergent point. This is because  the noise can affect the objective value at each feedback, which makes the updated $\{\Wb,\mPhi,\Cb\}$ fluctuate at the convergent point.

The sum MSE performance of different reflection elements for the proposed scheme and the transmission scheme with known CSI is evaluated in Fig. $\ref{n_element}$. We can see that the sum MSE decreases with the increase of the number reflecting elements on the RIS. Although there is a gap between the proposed scheme and the scheme with perfect CSI, the channel estimation is not required at the RIS and the BS, which makes our proposed scheme more practical. Moreover, it is shown that performance gain by increasing the number of reflecting elements is not obvious compared with the benchmark. This is because the increased dimension of the input variables makes the optimization more difficult with the limited number of feedback. Furthermore, we can see that the sum MSE of the proposed scheme with ten reflecting elements
is nearly the same as the sum MSE of the known CSI scheme with two elements. This reveals that the implementation of more reflecting elements without the RF-chains can achieve the same performance with the known CSI scheme.

\begin{figure}[t]
  \centering
  \includegraphics[width=5in]{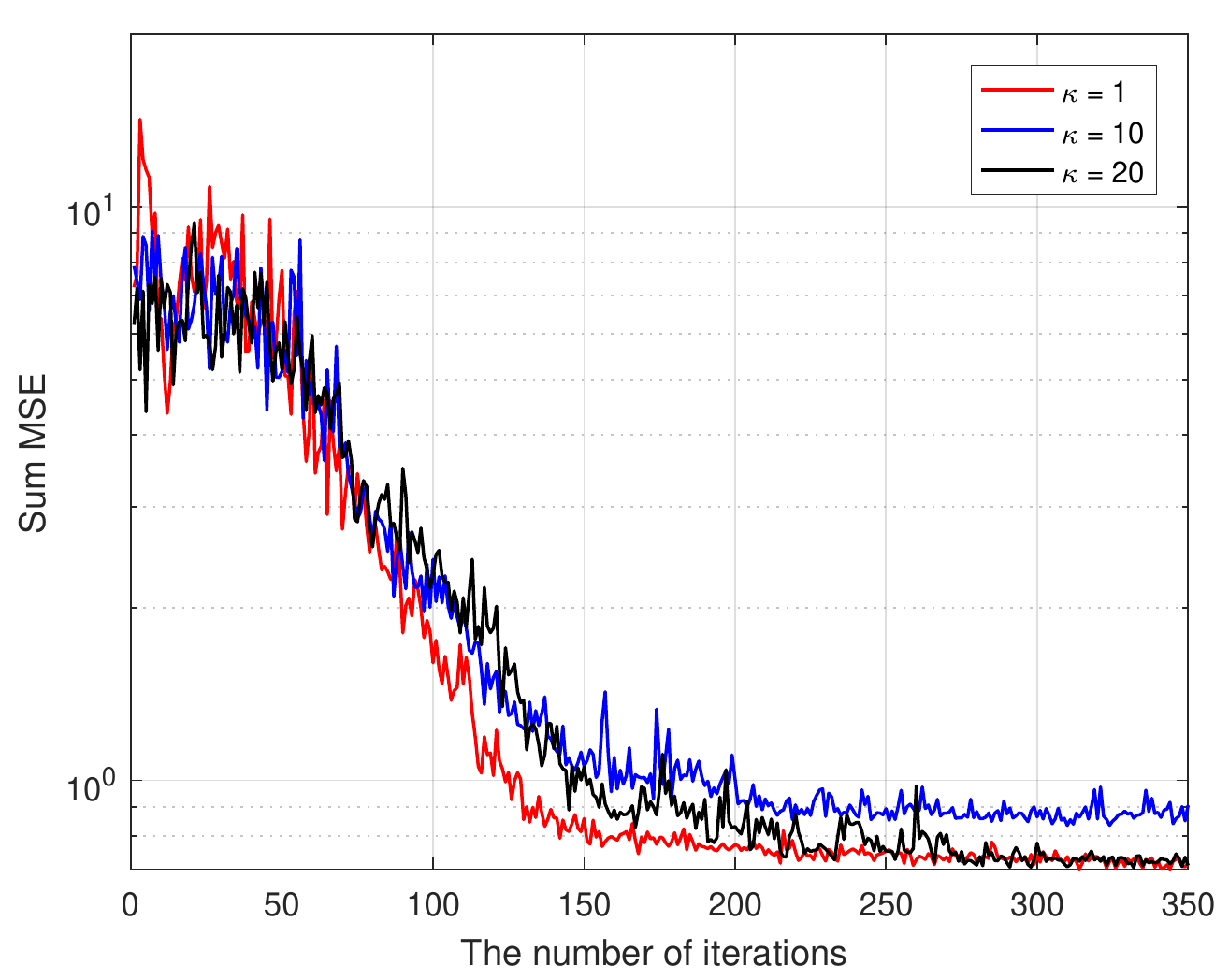} \\
  \caption{The sum MSE performance for the proposed transmission scheme with different number of pilot symbol vectors is evaluated in the downlink data transmission. We set $\mathsf{SNR} = 20$ dB. It is shown that the one pilot symbol vector is enough to the training process.}\label{miso_n_init}
\end{figure}

\begin{figure}[t]
  \centering
  \includegraphics[width=5in]{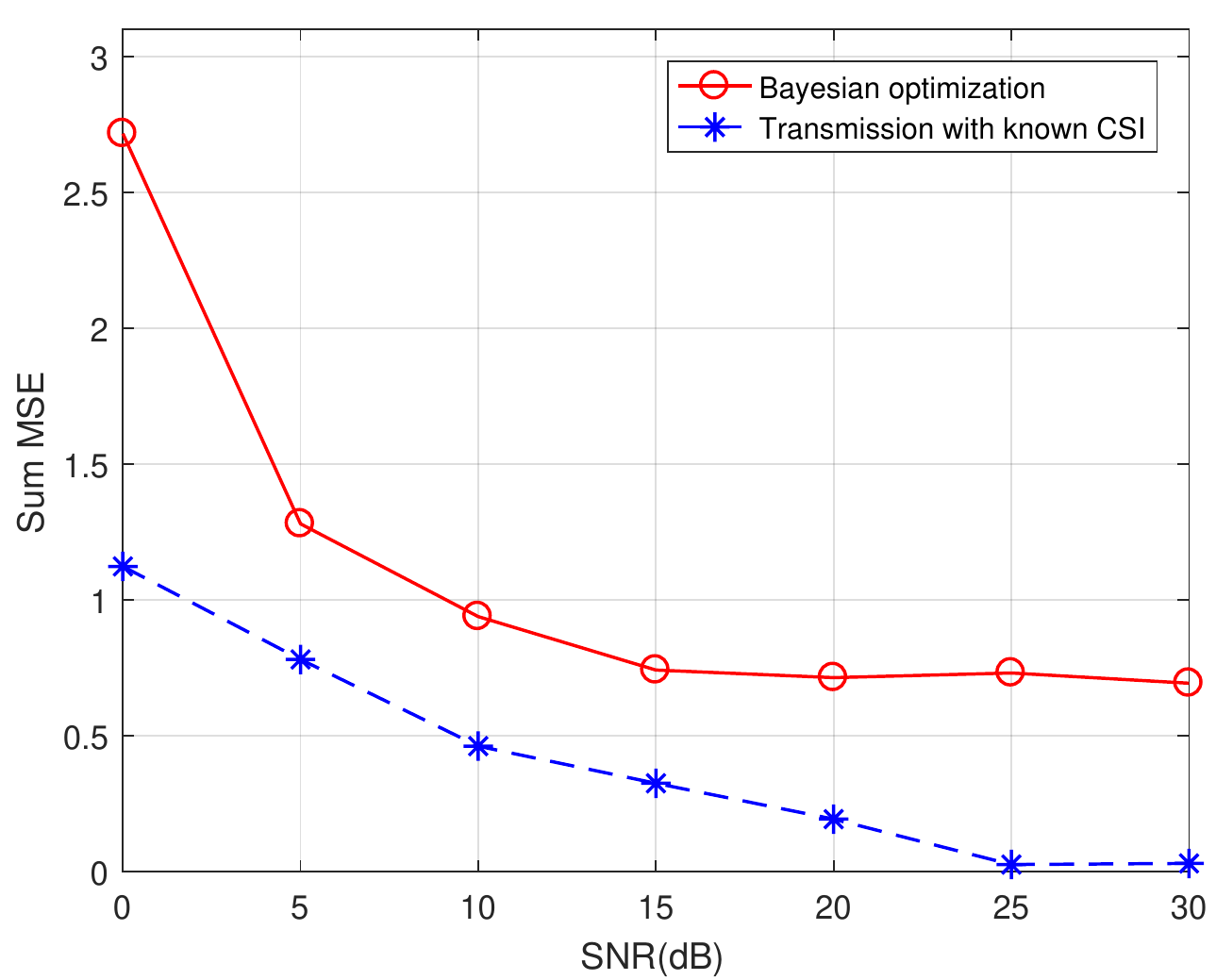} \\
  \caption{The sum MSE performance for the proposed transmission scheme with the slow fading channel is evaluated in the downlink data transmission. It is shown that the proposed ``Bayesian optimization'' is robust  to the slow varying channel. }\label{slowfading}
\end{figure}


The sum MSE performance with the different numbers of pilot symbol vectors is evaluated in Fig. \ref{miso_n_init}. The SNR is set to $20$ dB. We can see that the number of pilot symbol vectors does not significantly affect performance. Moreover, the sum MSE of the transmission with one pilot symbol vector outperforms the transmission with 10 and 20 pilot symbol vectors. This is because suitable noise can make the objective value avoid the local minimum and accelerate the convergence speed. Therefore, one pilot symbol vector is enough in the training process.

Next, we plot the MSE performance of the proposed scheme with the slow fading channel in Fig. \ref{slowfading}. We assume that $\Hb$ and $\Fb$ varies slowly in each feedback. The varying channel coefficients of each feedback are models as $\tilde \Hb =\Hb + \Delta\Hb $ and $\tilde \Fb =\Fb + \Delta\Fb $, where the entries of $\Delta\Hb, \Delta\Fb$ are $\in \mathcal{CN}(0,\nu^2)$. Here we set $\nu = 0.001$. We can see that the proposed bayesian optimization algorithm is robust to the slow fading channel. This is because the proposed scheme adopts the previous sample points and is not sensitive to the trivial parameter change. As a comparison, the frequent channel estimation is required for the known CSI scheme to avoid the performance loss with the channel varying.

\begin{figure}[t]
  \centering
  \includegraphics[width=5in]{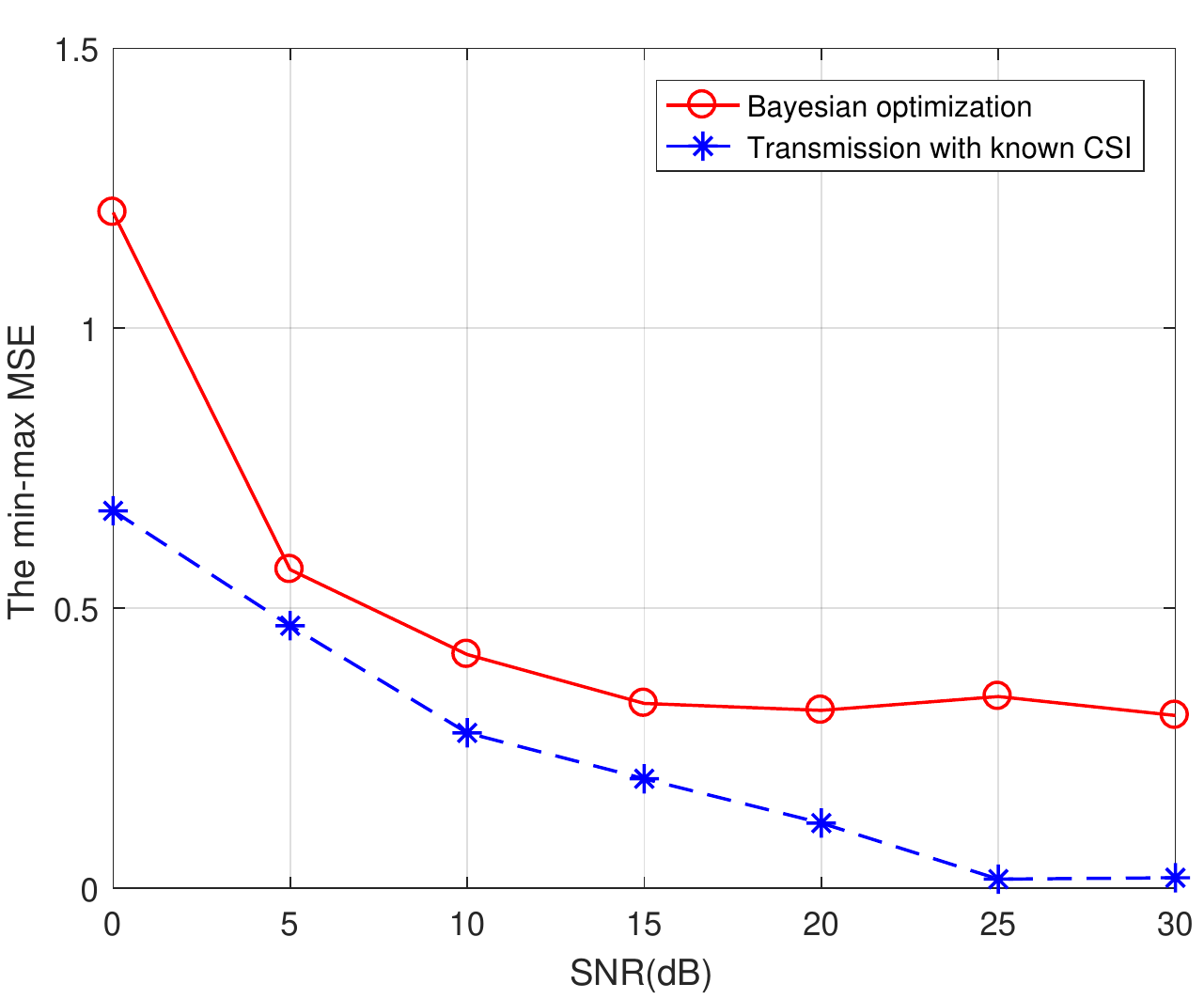} \\
  \caption{The MSE performance for the proposed scheme and the transmission scheme with known CSI is evaluated in the fairness downlink data transmission. The proposed ``Bayesian optimization'' achieves an acceptable performance with ``Transmission with known CSI''.}\label{minmax_mse}
\end{figure}

In Fig. \ref{minmax_mse}, The MSE performance for the proposed scheme and the transmission scheme with known CSI is evaluated in the fairness downlink data transmission. The known CSI scheme is the benchmark for the fairness problem, and the derivation of the known CSI scheme is similar to \cite{min-max_ris}.  As we can see, the min-max MSE of the proposed scheme achieves the acceptable performance compared with the known CSI scheme. Although the sum MSE problem and the min-max MSE problem are different in the formulation, the proposed scheme can solve these problems in a similar way.

\subsection{Downlink Power Transfer}
We further consider the large scale fading in the downlink power transfer. The large scale fading is modeled as $\kappa = \varsigma d_0^{-\alpha}$, where $d_0$ is the distance between the transmitter and the receiver; $\alpha= 2.2$ is the path loss exponent, and $\varsigma$ is the path loss at the reference distance $1$m which is set to $30$ dB. Moreover, the background noise variance at each user is set to $-110$ dBm. Moreover, the distance between the BS and the users is fixed to $50$ m, and the distance between the users and the RIS is set to $40$ m.

\begin{figure}[t]
  \centering
  \includegraphics[width=5in]{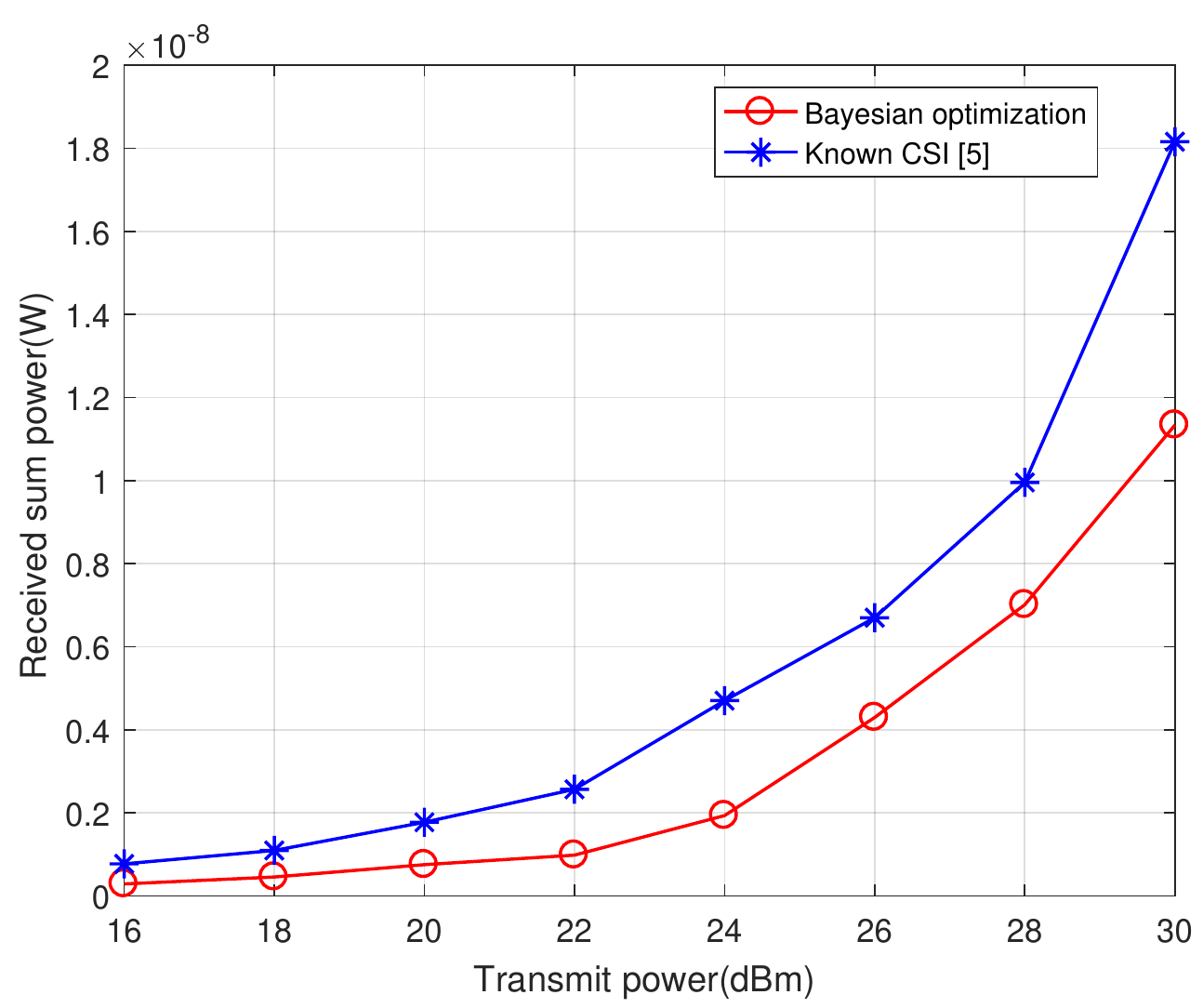} \\
  \caption{The received power performance with different transmit power for the proposed scheme and the transmission scheme with known CSI is evaluated in the downlink wireless power transfer. It is shown that the proposed ``Bayesian optimization'' achieves an acceptable performance with ``Known CSI'' in [5].}\label{sumP_vs_ptran}
\end{figure}

The received power performance for the proposed scheme and the transmission scheme with known CSI is evaluated in Fig. $\ref{sumP_vs_ptran}$. The known CSI scheme in [5] is set to the benchmark.  The received power of the users improves as the increase of the transmit power. The performance gap between the known CSI scheme and the proposed transmission scheme is acceptable.

\begin{figure}[t]
  \centering
  \includegraphics[width=5in]{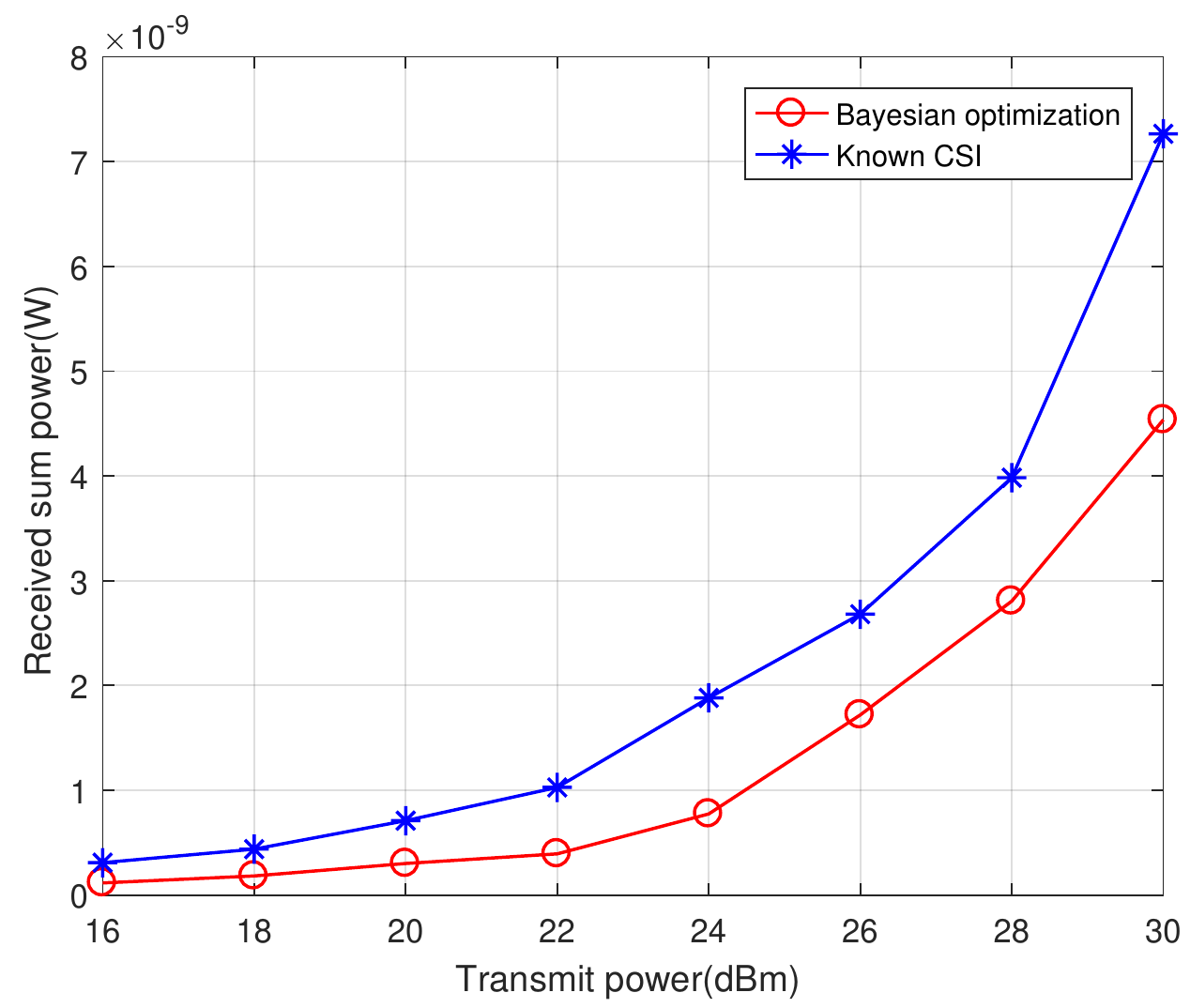} \\
  \caption{The received power performance with different transmit power for the proposed scheme and the transmission scheme with known CSI is evaluated in the downlink fairness wireless power transfer. The proposed ``Bayesian optimization'' has an acceptable performance with ``Known CSI''.}\label{sumP_vs_dist_maxmin}
\end{figure}

In Fig. \ref{sumP_vs_dist_maxmin}, the received power performance with different transmit power for the proposed scheme and the transmission scheme with known CSI is evaluated in the downlink fairness wireless power transfer. The derivation of the benchmark ``Known CSI" is similar to \cite{min-max_ris}. Our proposed Bayesian optimization based scheme achieves a satisfactory performance compared with the known CSI scheme. It validates the effectiveness of the proposed scheme in both the sum power maximization and the min-max problem.


\section{Conclusion}\label{S7}
In this paper, we studied the transmission scheme in the downlink RIS-assisted system with unknown CSI. The training scheme based on Bayesian optimization  was proposed to minimize the sum MSE. Moreover, the proposed scheme can be extended to the fairness problem and the wireless power transfer system. The superiority of the proposed scheme over the existing scheme is that the RF chains of the RIS and the channel estimation are not required for the transmission, which makes the implementation of the RIS more flexible and energy effective. The simulation results has demonstrated that the proposed scheme achieved an acceptable performance compared with the known CSI scheme. Furthermore, it was also shown that the proposed scheme can resist the impact of the slow fading channel. The ramification of this paper is that it provides a new perspective to design the RIS-assisted downlink transmission scheme.
\begin{appendix}
\section*{The Transmission Scheme With Known CSI}
In this section, we iteratively solve the original problem when CSI is available at the BS and the RIS. The closed-form solution is obtained to minimize the sum MSE. Lastly, the overall algorithm is provided.

We can expand the objection function \eqref{eq:maxuplink_multi} as
\begin{align}\label{eq:mmse_obj}
\mathcal{J}(\Wb, \mPhi,\Cb) &= \mathbb{E}\{\|(\Cb\Fb\mPhi\Hb\Wb-\Ib) \ssb + \Cb\ub\|^2\} \\
               &= \tr\{(\Cb\Fb\mPhi\Hb\Wb - \Ib)^{\rm H}(\Cb\Fb\mPhi\Hb\Wb - \Ib)\} + \sigma^2\tr\{\Cb^{\rm H}\Cb\} \\
               &= \tr\{\Wb^{\rm H}\Hb^{\rm H}\mPhi^{\rm H}\Fb^{\rm H}\Cb^{\rm H}\Cb\Fb\mPhi\Hb\Wb\}- \tr\{\Wb^{\rm H}\Hb^{\rm H}\mPhi^{\rm H}\Fb^{\rm H}\Cb^{\rm H}\} \notag\\
               &\qquad -\tr\{\Cb\Fb\mPhi\Hb\Wb\} + \sigma^2\tr\{\Cb^{\rm H}\Cb\}.
\end{align}
In the following, we focus on the optimization of the transmit beamformer $\Wb$ and the phase shift matrix $\mPhi$.
\subsection{Optimize $\{\Wb, \Cb\}$ for fixed $\mPhi$}
We refer to the lagrangian method to solve this problem. We introduce a introduce an auxiliary
variable $\alpha$. Then, let
\begin{align}
\Wb &= \alpha\bar\Wb, \\
\Cb &= \alpha^{-1}\bar\Cb,
\end{align}

and the lagrangian function can be written as
\begin{align}\label{eq:lag_obj}
\mathcal{L}(\bar\Wb,\alpha) &= \tr\{\bar\Wb^{\rm H}\Hb^{\rm H}\mPhi^{\rm H}\Fb^{\rm H}\bar\Cb^{\rm H}\bar\Cb\Fb\mPhi\Hb\bar\Wb\}- \tr\{\bar\Wb^{\rm H}\Hb^{\rm H}\mPhi^{\rm H}\Fb^{\rm H}\bar\Cb^{\rm H}\} \notag\\
              &\qquad -\tr\{\bar\Cb\Fb\mPhi\Hb\bar\Wb\} + \alpha^{-2}\sigma^2\tr\{\bar\Cb^{\rm H}\bar\Cb\}+\lambda(\alpha^2\tr\{\bar\Wb\bar\Wb^{\rm H}\}- P),
\end{align}
where $\lambda$ is the Lagrangian multiplier. Then, By setting $\frac{\partial \mathcal{L}}{\partial \lambda}= 0$ and $\frac{\partial \mathcal{L}}{\partial \alpha}= 0$. We can obtain
\begin{align}
\alpha^2 &=  \frac{P}{\tr\{\bar\Wb\bar\Wb^{\rm H}\}} \\
\alpha^4 &=  \frac{\sigma^2\tr\{\bar\Cb^{\rm H}\bar\Cb\}}{\lambda\tr\{\bar\Wb\bar\Wb^{\rm H}\}}  .
\end{align}
Hence, we have
\begin{equation}
\lambda\alpha^2 = \frac{\sigma^2}{P}\tr\{\bar\Cb^{\rm H}\bar\Cb\}.
\end{equation}
By setting $\frac{\partial \mathcal{L}}{\partial {\Wb^*}} = 0$, we can get the closed-form solution of $\bar\Wb$ and $\alpha$ as
\begin{align}
\bar\Wb &= 2\left(\Hb^{\rm H}\mPhi^{\rm H}\Fb^{\rm H}\bar\Cb^{\rm H}\bar\Cb\Fb\mPhi\Hb+ \frac{\sigma^2}{P}\tr\{\bar\Cb^{\rm H}\bar\Cb\}\Ib\right)^{-1}\bar\Cb\Fb\mPhi\Hb, \\
\alpha &= P^{\frac12}\tr\{\bar\Wb\bar\Wb^{\rm H}\}^{-\frac12},
\end{align}
Then, the optimal $\Wb$ can be expressed as
\begin{equation}\label{opt_w}
  \Wb^{\text{opt}} = \alpha\bar\Wb.
\end{equation}
By setting $\frac{\partial \mathcal{L}}{\partial {\Cb^*}} = 0$, we can obtain
\begin{equation}\label{opt_c_bar}
  \bar\Cb = \left(\Fb\mPhi\Hb\bar\Wb\bar\Wb^{\rm H}\Hb^{\rm H}\mPhi^{\rm H}\Fb^{\rm H}+\sigma^2\Ib\right)^{-1}\bar\Wb^{\rm H}\Hb^{\rm H}\mPhi^{\rm H}\Fb^{\rm H},
\end{equation}
Then, we can obtain the closed form solution of $\Cb$ as
\begin{equation}\label{opt c}
  \Cb^{\text{opt}} = \alpha^{-1}\bar\Cb.
\end{equation}

\subsection{Optimize $\mPhi$ for fixed \{\Wb, \Cb\}}
For fixed \{\Wb, \Cb\}, the optimization problem \eqref{opt_1} can be rewritten as
\begin{subequations}\label{opt_2_theta}
\begin{align}
\qquad	 \mathop{\min}_{\mPhi} ~~& \tr\{\mPhi^{\rm H}\Ab\mPhi\Bb\}-\tr\{\mPhi^{\rm H}\Dbb^{\rm H}\} - \tr\{\mPhi\Dbb\} \label{eq:maxuplink_multi_theta}\\
	\st    ~~&  0 \leq \theta_n \leq 2\pi, ~~\forall n= 1,2, \dots, N
\end{align}
\end{subequations}
where $\Ab =\Fb^{\rm H}\Cb^{\rm H}\Cb\Fb$, $\Bb = \Hb\Wb\Wb^{\rm H}\Hb^{\rm H}$ and $\Dbb = \Hb\Wb\Cb\Fb$. Since $\mPhi$ is the diagonal matrix, we have
\begin{align}\label{tr_fact}
  \tr\{\mPhi^{\rm H}\Ab\mPhi\Bb\} &= \mphi^{\rm H}(\Ab\odot\Bb)\mphi, \\
  \tr\{\mPhi\Dbb\} &=\db^{\rm T}\mphi, \\
  \tr\{\mPhi^{\rm H}\Dbb^{\rm H}\} &= \mphi^{\rm H}\db^*.
\end{align}
where $\db = \text{diag}\{\Dbb\}$. Then, the problem \eqref{opt_2_theta} can be rewritten as
\begin{subequations}\label{opt_2_theta_re}
\begin{align}
\qquad	 \mathop{\min}_{\mphi} ~~& \mphi^{\rm H}\mXi\mphi-\db^{\rm T}\mphi - \mphi^{\rm H}\db^*\label{eq:maxuplink_multi_theta_re}\\
	\st    ~~&   | \phi_n | = 1, ~~\forall n= 1,2, \dots, N,
\end{align}
\end{subequations}
where $\mXi = \Ab\odot\Bb$. 
Then, the problem \eqref{opt_2_theta_re} can be rewritten as
\begin{subequations}\label{eq:MM}
\begin{align}
\qquad	 \mathop{\min}_{\mphi} ~~& f(\mphi) \label{eq:MM_obj}\\
	\st    ~~&   | \phi_n | = 1, ~~\forall n= 1,2, \dots, N , \label{eq:molu_con}
\end{align}
\end{subequations}
where $f(\mphi) = \mphi^{\rm H}\mXi\mphi - 2 \Re\{\mphi^{\rm H}\db\}$. Since the unit modulus constraint in \eqref{eq:molu_con}, the problem \eqref{eq:MM} is a non-convex optimization problem and hard to solve. Here, we adopt the MM algorithm to solve this problem \cite{mutigroup_MISO}. The main idea is to construct a series of more tractable approximate subproblems. Similarly to the lemma in \cite{lemma}, we introduce the lemma in the following.
\begin{lemma}
For any given solution $\mphi^{t}$ at the $t$-th iteration and for any feasible $\mphi$, we have
\begin{equation}
\mphi^{\rm H}\mXi\mphi \leq y(\mphi|\mphi^t) \triangleq \mphi^{\rm H}\Xb\mphi - 2\Re\{\mphi^{\rm H}(\Xb-\mXi)\mphi\}+ (\mphi^{t})^{\rm H}(\Xb-\mXi)\mphi^{t}, \label{eq:lemma1}
\end{equation}
where $\mXi = \lambda_{\text{max}}\Ib_N$ and $\lambda_{\text{max}}$ is the maximum eigenvalue of $\mXi$. $\hfill\square$
\end{lemma}

Then, the surrogate objective function $y(\mphi|\mphi^t)$ can be constructed as
\begin{equation}\label{surro}
  g(\mphi|\mphi^t) = y(\mphi|\mphi^t) + 2 \Re\{\mphi^{\rm H}\db\}.
\end{equation}
It can be proofed that $g(\mphi|\mphi^t)$ satisfies the three conditions in \cite{mm}. Furthermore, the objective function $ g(\mphi|\mphi^t)$ is more tractable than the original $ f(\mphi|\mphi^t)$. Then, the subproblem can be solved at the $t$-th iteration is given by
\begin{subequations}\label{eq:g_MM}
\begin{align}
\qquad	 \mathop{\min}_{\mphi} ~~& g(\mphi|\mphi^t) \label{eq:g_obj}\\
	\st    ~~&   | \phi_n | = 1, ~~\forall n= 1,2, \dots, N.  \label{eq:g_molu_con}
\end{align}
\end{subequations}
\begin{algorithm}[t]
\caption{The transmission scheme with known CSI}
\label{alg:known_csi}
\begin{algorithmic}[1]
\State {Initial: $ t =1$, $\mphi = \mphi_0$, given ${\Hb}$, ${\Fb}$};
\While  {the $|f(\mphi^{t+1})-f(\mphi^{t})|$ is reduced by more than $\varepsilon$}
\State Calculate {$ \{\Wb$, $\Cb \}$} using \eqref{opt_w} and \eqref{opt c}, respectively;
\State Calculate $\qb^t$ using \eqref{eq:q};
\State Calculate $\mphi^{t+1}$ using \eqref{eq:phi_t};
\State Calculate $f(\mphi^{t+1})$ using \eqref{eq:MM_obj};
\EndWhile
\end{algorithmic}
\end{algorithm}
\end{appendix}
Since $\mphi^{\rm H}\mphi = M$, we have $\mphi^{\rm H}\mXi\mphi = M\lambda_{\text{max}}$. By substituting \eqref{eq:lemma1} into \eqref{eq:g_obj}, the problem \eqref{eq:g_MM} can be reformulated as
\begin{subequations}\label{eq:max_g_MM}
\begin{align}
\qquad	 \mathop{\max}_{\mphi} ~~& 2 \Re\{\mphi^{\rm H}\qb^t\} \label{eq:g_obj}\\
	\st    ~~&   | \phi_n | = 1, ~~\forall n= 1,2, \dots, N.  \label{eq:g_molu_con}
\end{align}
\end{subequations}
where
\begin{equation}\label{eq:q}
  \qb = (\lambda_{\text{max}}\Ib_N - \mXi)\mphi^t-\db^*.
\end{equation}
It can be easily seen that the optimal solution of problem \eqref{eq:max_g_MM} is given by
\begin{equation}\label{eq:phi_t}
  \phi^{t+1} = e^{j\arg(\qb^t)}.
\end{equation}
Finally, we provide the transmission scheme with known CSI in Algorithm $\ref{alg:known_csi}$.

\bibliographystyle{IEEEtran}

\bibliography{Reference}
\end{document}